\documentclass[aps,prd,preprintnumbers,superscriptaddress,nofootinbib,showpacs,twocolumn]{revtex4}%
\usepackage[dvipdfmx]{graphicx}
\usepackage{bm,latexsym,amsmath,amssymb,amsfonts,mathrsfs}
\usepackage{color}
\input{colordvi.tex}
\usepackage[dvipdfmx]{hyperref}
\usepackage{url}
\hypersetup{
    colorlinks=true,
    citecolor=cyan,
}
\newcommand*{\D}{{\rm d}}
\newcommand*{\mpl}{M_{\rm Pl}}
\begin{document}

\title{Scale-invariant perturbations from NEC violation: A new variant of Galilean Genesis}

\author{Sakine~Nishi}
\email[Email: ]{sakine"at"rikkyo.ac.jp}
\affiliation{Department of Physics, Rikkyo University, Toshima, Tokyo 171-8501, Japan}

\author{Tsutomu~Kobayashi}
\email[Email: ]{tsutomu"at"rikkyo.ac.jp}
\affiliation{Department of Physics, Rikkyo University, Toshima, Tokyo 171-8501, Japan}

\begin{abstract}
We propose a novel branch of the Galilean Genesis scenario as an alternative to inflation,
in which the universe starts expanding from Minkowski in the asymptotic past with a gross
violation of the null energy condition (NEC).
This variant, described by several functions and parameters within the Horndeski scalar-tensor theory,
shares the same background dynamics with the existing Genesis models,
but the nature of primordial quantum fluctuations is quite distinct.
In some cases, tensor perturbations grow on superhorizon scales.
The tensor power spectrum can be red, blue, or scale invariant, depending on the model,
while scalar perturbations are nearly scale invariant.
This is in sharp contrast to typical NEC-violating cosmologies, in which a blue tensor tilt is generated.
Though the primordial tensor and scalar spectra are both nearly scale invariant
as in the inflationary scenario, the consistency relation in our variant of Galilean Genesis
is non-standard.
\end{abstract}

\pacs{
98.80.Cq, 
04.50.Kd  
}
\preprint{RUP-16-28}
\maketitle

\section{Introduction}


It is no exaggeration to say that inflation~\cite{r2, inflation1, inflation2} is now
a part of the ``standard model'' of the Universe.
Not only homogeneity, isotropy, and flatness of space,
but also the inhomogeneous
structure of the Universe originated from tiny primordial fluctuations~\cite{Mukhanov:1981xt},
can be elegantly explained by a phase of quasi-de Sitter expansion in the early Universe.
However, even the inflationary scenario cannot resolve the initial singularity problem~\cite{Borde:1996pt},
which raises the motivation for debating the possibilities of alternatives to
inflation~(for a review, see, e.g., Refs.~\cite{Brandenberger:2009jq, Brandenberger:2016vhg}).
In order to be convinced that the epoch of quasi-de Sitter expansion did exist in the early Universe,
one must rule out such alternatives.

A typical feature of singularity-free alternative scenarios is
that the Hubble parameter $H$ is an {\em increasing} function of time in the early universe.
The null energy condition requires that for all null vectors $k^\mu$
the energy-momentum tensor satisfies $T_{\mu\nu}k^\mu k^\nu\ge 0$,
which, upon using the Einstein equations, translates to the condition for
the Ricci tensor, $R_{\mu\nu}k^\mu k^\nu\ge 0$.
In a cosmological setup this reads $\dot H\le 0$,
and hence the NEC\footnote{In this paper, we use the terminology NEC
when referring to $R_{\mu\nu}k^\mu k^\nu\ge 0$, which is, more properly, the null convergence condition.}
is violated in such alternative scenarios.
Unfortunately, in many cases the violation of the NEC implies
that the system under consideration is unstable.
Earlier NEC-violating models are indeed precluded by this instability issue~\cite{Battefeld:2014uga}.
Recently, however, it was noticed that
scalar-field theories with second-derivative Lagrangians
admit stable NEC-violating solutions~\cite{Creminelli:2010ba,Deffayet:2010qz,Kobayashi:2010cm},
which revitalizes singularity-free alternatives to
inflation~\cite{Qiu:2011cy,Easson:2011zy,Cai:2012va,Cai:2013vm,Osipov:2013ssa,Qiu:2013eoa,Liu:2014tda,Rubakov:2014jja}. 
One can avoid the initial singularity also in emergent universe cosmology~\cite{Ellis:2002we, Ellis:2003qz, Cai:2012yf, Cai:2013rna} and
in string gas cosmology~\cite{Brandenberger:1988aj, Battefeld:2005av}.

The future
detection of primordial gravitational waves
(tensor perturbations)
is supposed to give us valuable information of the early Universe.
It is folklore that a
nearly scale-invariant red spectrum of primordial gravitational waves
is the ``smoking gun'' of inflation.
The reason that this is believed to be so is the following.
The amplitude of each gravitational wave mode
is determined solely by the value of the Hubble parameter
evaluated at horizon crossing.
During inflation $H$ is a slowly decreasing function of time,
while in alternative scenarios the time evolution of $H$ is very different.
This folklore is not true, however, even in the context of inflation,
because some extended models of inflation
can violate the NEC stably and
thereby the Hubble parameter slowly increases,
giving rise to nearly scale-invariant {\em blue} tensor spectra~\cite{Kobayashi:2010cm}.
Then, does the detection of nearly scale-invariant tensor perturbations
indicate a phase of quasi-de Sitter expansion?
Naively, the gross violation of the NEC
in alternative models implies strongly blue tensor spectra,
and by this feature one would be able to discriminate inflation from alternatives.
In this paper, we show that this expectation is not true:
nearly scale-invariant scalar and tensor perturbations
can be generated from quantum fluctuations on a NEC-violating background.\footnote{It has been known that
in string gas cosmology scale-invariant scalar and tensor perturbations are generated
from thermal string fluctuations~\cite{Brandenberger:2009jq,Brandenberger:2016vhg}.
Nearly scale-invariant tensor perturbations can also be sourced by
gauge fields in bouncing~\cite{Ben-Dayan:2016iks} and ekpyrotic~\cite{Ito:2016fqp} scenarios.}
Thus, it is possible that the individual spectrum has no difference from that of inflation,
though the consistency relation turns out to be different.

The model we present in this paper is
a variant of Galilean Genesis~\cite{Creminelli:2010ba},
in which the universe starts expanding from Minkowski by violating the NEC stably.
The earlier proposal of Galilean Genesis~\cite{Creminelli:2010ba, Creminelli:2012my, Hinterbichler:2012fr, Hinterbichler:2012yn}
fails to produce
scale-invariant curvature perturbations (without invoking the curvaton), but
it was shown in~\cite{Nishi:2015pta,Liu:2011ns,Piao:2010bi} that
it is possible if one generalizes the original models.
In all those models, the tensor perturbations have strongly blue spectra
and hence the amplitudes are too small to be detected at low frequencies~\cite{Nishi:2016wty}.
In our new models of Galilean Genesis, however, the primordial tensor spectrum
can be red, blue, or scale invariant, depending on the parameters of the model,
and the curvature perturbation can have a nearly scale-invariant spectrum.
We work in the Horndeski theory~\cite{Horndeski, Deffayet:2011gz, Kobayashi:2011nu},
the most general scalar-tensor theory
with second-order field equations, to construct a general Lagrangian
admitting the new Genesis solution with the above-mentioned properties.
As a specific case our Lagrangian includes
the Genesis model recently obtained by Cai and Piao~\cite{Cai:2016gjd},
which yields scale-invariant tensor perturbations and {\em strongly red} scalar perturbations.

The plan of this paper is as follows.
In Sec.~II, we introduce the general Lagrangian for our new variant of Galilean Genesis,
and study the background evolution to discuss whether
homogeneity, isotropy, and flatness of space can be explained in the present scenario.
Then, in Sec.~III, we calculate primordial scalar and tensor spectra.
We give a concrete example yielding scale-invariant scalar and tensor perturbations in Sec.~IV.
In Sec.~V we draw our conclusions.

\section{A new variant of Generalized Galilean Genesis}

We work in the
Horndeski theory (also known as the Generalized Galileon theory)~\cite{Horndeski, Deffayet:2011gz, Kobayashi:2011nu},
whose action is given by
\begin{eqnarray}
S&=&\int\D^4x\sqrt{-g}\left({\cal L}_2+{\cal L}_3+{\cal L}_4+{\cal L}_5\right),\label{Hor}
\end{eqnarray}
with
\begin{eqnarray}
&&{\cal L}_2=G_2(\phi, X),\nonumber \\
&&{\cal L}_3=-G_3(\phi, X)\Box\phi,\nonumber\\
&&{\cal L}_4=G_4(\phi, X)R+G_{4X}\left[(\Box\phi)^2-(\nabla_\mu\nabla_\nu\phi)^2\right],\nonumber\\
&&{\cal L}_5=G_5(\phi, X)G^{\mu\nu}\nabla_\mu\nabla_\nu\phi-\frac{1}{6}G_{5X}\bigl[(\Box\phi)^3\nonumber\\
&&\qquad\qquad -3\Box\phi(\nabla_\mu\nabla_\nu\phi)^2+2(\nabla_\mu\nabla_\nu\phi)^3\bigr],
\end{eqnarray}
where $X$ is the kinetic term of the scalar field $\phi$,
\begin{eqnarray}
X:=-\frac{1}{2}g^{\mu\nu}\partial_\mu\phi\partial_\nu\phi,
\end{eqnarray}
and $G_{i}(\phi,X)\, (i=2,3,4,5)$ are arbitrary functions of $\phi$ and $X$.
The subscript $X$ stands for differentiation with respect to $X$.

Let us begin with a brief review on (generalized) Galilean Genesis.
In a previous paper~\cite{Nishi:2015pta} we developed a unifying framework
for the Genesis scenarios in which the universe starts expanding
from Minkowski in the asymptotic past.
The framework is based on the following choice of the Horndeski functions:
\begin{align}
& G_2=e^{2(\alpha+1)\lambda\phi}g_2(Y),\quad
&&G_3=e^{2\alpha\lambda\phi}g_3(Y),
\notag \\
&G_4=\frac{\mpl^2}{2}+e^{2\alpha\lambda\phi}g_4(Y),
\quad
&&G_5=e^{-2\lambda\phi}g_5(Y),
\end{align}
where each $g_i$ is an arbitrary function of
\begin{align}
Y:=e^{-2\lambda\phi}X,
\end{align}
$\alpha$ is a parameter, and
$\lambda$ is introduced so that $\phi$ has the dimension of mass.
The field equations admit the cosmological solution of the form
\begin{align}
Y\simeq Y_0={\rm const}\;\;\Rightarrow&\;\;e^{\lambda \phi}\simeq\frac{1}{\lambda\sqrt{2Y_0}}\frac{1}{(-t)},
\\
H\simeq \frac{h_0}{(-t)^{2\alpha + 1}}
\;\;\Rightarrow&\;\;
a\simeq 1+\frac{1}{2\alpha}\frac{h_0}{(-t)^{2\alpha}},
\end{align}
for large $|t|$, $(-t)^{2\alpha}\gg h_0$.
This solution describes the universe emerging from Minkowski,
and its expansion rate is controlled by the parameter $\alpha$.
The above generalized Galilean Genesis framework
can reproduce different concrete
models~\cite{Creminelli:2010ba, Creminelli:2012my, Hinterbichler:2012fr, Hinterbichler:2012yn,Liu:2011ns,Piao:2010bi}
as specific cases by 
choosing $\alpha$ and the forms of $g_i(Y)$.

The evolution of the curvature perturbation $\zeta$ in generalized Galilean Genesis
is intriguing, as $\zeta$ {\em grows} even on superhorizon scales for $\alpha > 1/2$.
This fact was first found in the original Galilean Genesis model~\cite{Creminelli:2010ba}
which corresponds to $\alpha = 1$.
Interestingly, the spectral index is completely determined by the parameter
as
\begin{align}
n_s=5-2\alpha
\end{align}
(in the $\alpha>1/2$ case), and therefore we have the scale-invariant
curvature perturbations for $\alpha=2$.
The superhorizon growth of $\zeta$ during the Genesis phase is
analogous to that in the so-called non-attractor inflation models~\cite{Kinney:2005vj,Inoue:2001zt}
and in bounce models~\cite{Finelli:2001sr,Wands:1998yp,Allen:2004vz}.
In contrast to the curvature perturbation,
the tensor perturbations
feel very slow cosmic expansion and so
are living effectively in Minkowski.
This results in a blue-tilted spectrum irrespective of $\alpha$ and $g_i(Y)$~\cite{Nishi:2015pta,Nishi:2016wty}.

\subsection{A new Lagrangian for Galilean Genesis}

Now let us present a new variant of generalized Galilean Genesis
that enjoys a similar background evolution but exhibits
a novel behavior of perturbations compared to the existing Genesis models.
As the arbitrary functions $G_{i}(\phi,X)$ in the Horndeski theory we choose
\begin{eqnarray}
&& G_2=e^{2(\alpha+1)\lambda\phi}g_2(Y) \nonumber \\
&& \quad  \quad\;\; +e^{-2(\beta-1)\lambda\phi}a_2(Y) +e^{-2(\alpha+2\beta-1)}b_2(Y), \nonumber  \\
&& G_3=e^{2\alpha\lambda\phi}g_3(Y)  \nonumber \\
&& \quad \quad \;\;+e^{-2\beta\lambda\phi}a_3(Y) +e^{-2(\alpha+2\beta)}b_3(Y), \nonumber\\
&& G_4=e^{-2\beta\lambda\phi}a_4(Y)+e^{-2(\alpha+2\beta)\lambda\phi}b_4(Y), \nonumber\\
&& G_5=e^{-2(\alpha+2\beta+1)\lambda\phi}b_5(Y), \label{func_G_adm}
\end{eqnarray}
where $g_2$ and $g_3$ are arbitrary functions of $Y$,
but $a_i(Y)$ and $b_i(Y)$ are such that
\begin{eqnarray}
&& a_2(Y)= 8\lambda^2Y (Y\partial_Y+\beta)^2 A(Y), \\
&& a_3(Y)= -2\lambda (2Y\partial_Y+1)(Y\partial_Y+\beta)A(Y), \\
&& a_4(Y)= Y\partial_Y A(Y), \\
&& b_2(Y)= 16\lambda^3 Y^2(Y\partial_Y+\alpha+2\beta+1)^3B(Y), \\
&& b_3(Y)= -4\lambda^2 Y(2Y\partial_Y+3)\nonumber \\
&& \quad \quad\quad \;\;\;\times(Y\partial_Y+\alpha+2\beta+1)^2B(Y), \quad\quad\quad \\
&& b_4(Y)= 2\lambda Y (Y\partial_Y+1) \nonumber \\
&& \quad \quad\quad\;\;\; \times(Y\partial_Y+\alpha+2\beta+1)B(Y), \\
&& b_5(Y)=-(2Y\partial_Y+1)(Y\partial_Y+1)B(Y),
\end{eqnarray}
with arbitrary functions $A(Y)$ and $B(Y)$. We thus have four functional degrees of freedom,
as well as two constant parameters $\alpha$ and $\beta$ in this setup.
We assume that
\begin{align}
\alpha+ \beta>0
\end{align}
in order to obtain the background evolution which we will present shortly.
However, at this stage we do {\em not} impose that $\alpha >0$ and $\beta>0$.

We assume the ansatz,
\begin{align}
Y\simeq Y_0={\rm const},\quad H\simeq \frac{h_0}{(-t)^{2\alpha+2\beta+1}}, \label{gen:back}
\end{align}
and substitute this into the field equations to
see that Eq.~(\ref{gen:back}) indeed gives a consistent solution for a large $|t|$.
(The range of $t$ is $-\infty<t<0$.) The scale factor for a large $|t|$ is given by
\begin{align}
a\simeq 1+\frac{1}{2(\alpha+\beta)}\frac{h_0}{(-t)^{2(\alpha+\beta)}}. \label{gen:back_a}
\end{align}
The (00) and ($ij$) components of the gravitational field equations read, respectively,
\begin{align}
\hat\rho(Y_0)+{\cal O}(|t|^{-2(\alpha+\beta)}) &=0,\label{gen:eq1}
\\
2{\cal G}_T\dot H + e^{2(\alpha +1)\lambda\phi}\hat p(Y_0)+{\cal O}(|t|^{-2(2\alpha+\beta+1)})&=0,\label{gen:eq2}
\end{align}
where we defined
\begin{align}
\hat\rho(Y)&:=2Y g_2'-g_2-4\lambda Y\left(\alpha g_3-Yg_3'\right),
\\
\hat p(Y)&:=g_2-4\alpha \lambda Yg_3 ,
\end{align}
and
\begin{align}
{\cal G}_T&:=2\left[G_4-2XG_{4X}-X\left(H\dot\phi G_{5X} -G_{5\phi}\right)\right]\notag \\
&\simeq -2e^{-2\beta\lambda\phi}Y_0(A'+2YA'')\notag\\
&\quad +2 e^{-2(\alpha+2\beta+1)\lambda\phi}H\dot\phi Y_0(6B'+9YB''+2Y^2B'''). \label{defGT}
\end{align}
Note that
\begin{align}
{\cal G}_T\propto (-t)^{2\beta},\label{behaviorGT}
\end{align}
and hence ${\cal G}_T\dot H={\cal O}(|t|^{-2(\alpha+1)})$.
Equation~(\ref{gen:eq1})
fixes $Y_0$ as a root of 
\begin{align}
\hat\rho(Y_0)=0,\label{co1}
\end{align}
and then Eq.~(\ref{gen:eq2}) is used to determine $h_0$.
Since there is $H$ in ${\cal G}_T$, Eq.~(\ref{gen:eq2})
reduces to a quadratic equation in $h_0$ in general.
We have sensible NEC-violating cosmology only for $h_0>0$.
Since it will turn out that the condition
\begin{eqnarray}
{\cal G}_T>0
\end{eqnarray}
is required from the stability of tensor perturbations, one must impose
\begin{eqnarray}
\hat{p}(Y_0)<0 ,
\end{eqnarray}
though this is not a sufficient condition for $h_0>0$.

A particular case of this class of Genesis models can be found in~\cite{Cai:2016gjd},
which corresponds to $A\propto Y^{-2}$, $B=0$ with $\alpha = \beta=2$.

We have thus found that the Horndeski theory with~(\ref{func_G_adm})
admits the Genesis solution~(\ref{gen:back}),
which is similar to previous ones~\cite{Nishi:2015pta}.
However, we will show in the next section that
the evolution of tensor perturbations is quite different:
they can even grow on superhorizon scales and
can give rise to a variety of values of the spectral index $n_t$.
Before seeing this, let us address more about the background evolution.

\subsection{Flatness Problem}

Now let us move on to discuss the problems which inflation solves.
In the inflationary universe, the curvature term in the Friedmann equation
is diluted exponentially relative to the other terms, and thus the flatness problem
in standard Big Bang cosmology is resolved.
Since cosmic expansion is very slow in Galilean Genesis, $a\simeq1$,
one may wonder if the flatness problem is solved as well in this scenario.
We have shown in the previous paper~\cite{Nishi:2015pta} that
the curvature term is eventually diluted away in all existing Galilean Genesis models.
We now check this point in our new variant of Galilean Genesis.

The background equations in the presence of the spatial curvature $K$ are given by~\cite{Nishi:2015pta}
\begin{align}
e^{2(\alpha+1)\lambda\phi}\hat\rho(Y_0)-\frac{3{\cal G}_T K}{a^2}&\simeq 0, \label{keq1}
\\
2{\cal G}_T\dot H + e^{2(\alpha +1)\lambda\phi}\hat p(Y_0)
+\frac{{\cal F}_T K}{a^2}&\simeq 0,\label{keq2}
\end{align}
where
\begin{align}
{\cal F}_T&:=2\left[G_4-X\left(\ddot\phi G_{5X}+G_{5\phi}\right)\right] \notag \\
&\simeq 2e^{-2\beta\lambda\phi}Y_0A'\nonumber\\
& \quad -4 e^{-2(\alpha+2\beta)\lambda\phi}(1+2\alpha+4\beta)\lambda Y_0^2(2B'+YB'').
\label{defFT}
\end{align}
We have
\begin{align}
{\cal F}_T\propto
\begin{cases}
    (-t)^{2\beta} & (2B'+Y_0B''=0) \\
    (-t)^{2(\alpha+2\beta)} & (2B'+Y_0B''\neq 0)
\end{cases}.\label{behaviorFT}
\end{align}
In order for the flatness problem to be resolved,
the curvature term has to be negligible compared to the other terms.
In Eq.~(\ref{keq1}) the ratio of the curvature term to the first term is
$\sim (-t)^{2(\alpha+\beta) + 1}$, and due to the condition $\alpha+\beta>0$
the curvature term becomes negligible as time proceeds.
It can be seen using Eq.~(\ref{behaviorFT}) that the same is true in Eq.~(\ref{keq2}).
We have thus confirmed that the flatness problem can be solved as well in our
new variant of Galilean Genesis.

\subsection{Anisotropy}\label{subsec:aniso}

Given the large-scale isotropy of the Universe,
let us consider the evolution of anisotropies in the Genesis phase and check
whether the universe can safely be isotropized.
In standard cosmology, the shear term in the Friedmann equation
is diluted rapidly as $\propto a^{-6}$.
However, the situation is subtle in Galilean Genesis.

We describe an anisotropic universe using the Kasner-type metric as
\begin{eqnarray}
\D s^2=-\D t^2+a^2\left[e^{2\theta_1(t)}\D x^2+e^{2\theta_2(t)}\D y^2+e^{2\theta_3(t)}\D z^2\right],~
\end{eqnarray}
where we define
\begin{eqnarray}
\theta_1=\beta_++\sqrt{3}\beta_-,\quad \theta_2=\beta_+-\sqrt{3}\beta_-, \quad \theta_3=-2\beta_+ .
\end{eqnarray}
From the equations of motion for $\beta_+$ and $\beta_-$ with $a\simeq 1$, we obtain~\cite{Nishi:2015pta}
\begin{eqnarray}
\frac{\D}{\D t}\left[{\cal G}_T \dot\beta_+-2X\dot\phi G_{5X}\left(\dot\beta_+^2-\dot\beta_-^2\right)
\right]&=&0, \label{anisoeq1}\\
\frac{\D}{\D t}\left[{\cal G}_T\dot\beta_-+4X\dot\phi G_{5X}\dot\beta_+ \dot\beta_-
\right]&=&0\label{anisoeq2},
\end{eqnarray}
where ${\cal G}_T\propto (-t)^{2\beta}$ and
\begin{align}
X\dot\phi G_{5X}&=\dot\phi e^{-2(\alpha+2\beta+1)\lambda\phi}Y_0 b_5'(Y_0)
\notag \\
&
\propto (-t)^{2\alpha + 4\beta+1}.
\end{align}

In the $b_5'(Y_0)=0$ case,
it is easy to see that $\dot \beta_\pm \propto (-t)^{-2\beta}$,
and hence
\begin{align}
\frac{\dot\beta_\pm}{H}\propto (-t)^{2\alpha + 1}.
\end{align}
This implies that if $\alpha>-1/2$,
the universe is isotropized as it expands.

To see what happens in the general case of $b_5'(Y_0)\neq 0$,
it is convenient to define
\begin{align}
b:=\frac{{\cal G}_T}{2X\dot\phi G_{5X}}\sim H\propto (-t)^{-2(\alpha+\beta)-1}.
\end{align}
One can integrate Eqs.~(\ref{anisoeq1}) and~(\ref{anisoeq2}) to obtain
\begin{align}
\left(\frac{\dot\beta_+}{b}\right)-\left(\frac{\dot\beta_+}{b}\right)^2
+\left(\frac{\dot\beta_-}{b}\right)^2
&={\rm const}\times (-t)^{2\alpha+1},\label{bbbeq1}
\\
\left(\frac{\dot\beta_-}{b}\right)+2\left(\frac{\dot\beta_+}{b}\right)\left(\frac{\dot\beta_-}{b}\right)
&={\rm const}\times (-t)^{2\alpha+1}.\label{bbbeq2}
\end{align}
If $\alpha < -1/2$, the right hand sides grow as the universe expands,
leading to the growth of $\dot\beta_\pm/b$, i.e., the growth of $\dot\beta_\pm/H$.
Therefore, this case is not acceptable.
If $\alpha > -1/2$ and the initial anisotropies are sufficiently smaller than $b\,(\sim H)$,
the quadratic terms in Eqs.~(\ref{bbbeq1}) and~(\ref{bbbeq2}) can be ignored
and we have $\dot\beta_\pm/b\propto (-t)^{2\alpha + 1}$, i.e.,
\begin{align}
\frac{\dot\beta_\pm}{H}\propto(-t)^{2\alpha + 1},
\end{align}
implying that the universe is isotropized.
However, if $\alpha>-1/2$ and the initial anisotropies are as large as $\dot\beta_\pm={\cal O}(b)$,
it is possible that the solution approaches one of the following attractors:
\begin{eqnarray}
(\dot\beta_+,\dot\beta_-)=(b,0),(-\frac{1}{2}b,\frac{\sqrt{3}}{2}b),(-\frac{1}{2}b,-\frac{\sqrt{3}}{2}b),
\end{eqnarray}
though a full phase-space analysis is beyond the scope of the paper.
In this case, the anisotropies remain,
\begin{align}
\frac{\dot\beta_\pm}{H}={\rm const},
\end{align}
which is not acceptable.

In light of the above result, it is required that
\begin{align}
\alpha >-\frac{1}{2}
\end{align}
to avoid a highly anisotropic universe.
We also assume in the general case of $b_5(Y_0)\neq 0$ that
the initial anisotropies $\dot\beta_\pm$ are not as large as $b\,(\sim H)$.

\section{Power Spectra}

In the previous section we have seen that
the Horndeski theory with~(\ref{func_G_adm})
offers a variant of generalized Galilean Genesis,
which has a similar background solution to the previous Genesis models.
A significant point of this variant is
found in the dynamics of scalar and tensor perturbations. In this section,
let us discuss the evolution of the cosmological perturbations and
their power spectra.
As we will see, the quadratic Lagrangian for the curvature perturbation $\zeta$ is of the form
\begin{align}
{\cal L}\sim C_1(-t)^{2p}\dot\zeta^2-C_2(-t)^{2q}(\Vec{\nabla}\zeta)^2,
\end{align}
where $C_1$, $C_2$, $p$, and $q$ are constants satisfying $C_1,\,C_2>0$ and $1-p+q>0$.
The Lagrangian for the tensor perturbations $h_{ij}$ is of the same form.
In Appendix~\ref{appendix1} we summarize the useful formulas for the power spectra
obtained from such quadratic Lagrangians, which we will refer to in the following discussion.

\subsection{Tensor Perturbations}\label{GWs}

The quadratic action for tensor perturbations in the Horndeski theory
is given in general by~\cite{Kobayashi:2011nu}
\begin{eqnarray}
S_h^{(2)}=\frac{1}{8}\int \D t\D^3x
\,a^3\left[{\cal G}_T\dot h^2_{ij}-\frac{{\cal F}_T}{a^2}(\Vec{\nabla} h_{ij})^2\right],
\label{action_hij}
\end{eqnarray}
where
${\cal G}_T$ and ${\cal F}_T$ were already defined in Eqs.~(\ref{defGT}) and~(\ref{defFT}).
In the previous models of generalized Galilean Genesis, as well as
in conventional models of inflation (in Einstein gravity), we have
${\cal G}_T,\,{\cal F}_T\simeq\,$const, giving
\begin{align}
h_{ij}\sim{\rm const}\;\;\;{\rm and}\;\;\;{\rm decaying~solution},
\end{align}
on superhorizon scales. The amplitude of the dominant constant mode
is proportional to $H$ at horizon crossing, leading to
a slightly red spectrum in the case of inflation and
a strongly blue spectrum in NEC-violating cosmologies such as Galilean Genesis.

In our new models of Galilean Genesis,
we still have $a\simeq 1$. However, now ${\cal G}_T$ and ${\cal F}_T$
are strongly time-dependent, as shown in Eqs.~(\ref{behaviorGT}) and~(\ref{behaviorFT}).
Noting that ${\cal G}_T\propto (-t)^{2\beta}$, we obtain
two independent solutions on superhorizon scales,
\begin{align}
h_{ij}\sim{\rm const}\;\;\;{\rm and}\;\;\;
\int^t\frac{\D t'}{a^3{\cal G}_T}\sim (-t)^{1-2\beta}.
\end{align}
This indicates that, while we have constant and decaying solutions as usual for $\beta<1/2$,
for $\beta > 1/2$ the would-be decaying mode {\em grows} on superhorizon scales.
This is in sharp contrast to the previous Genesis models.

This peculiar evolution of the tensor perturbations for $\beta>1/2$ can be explained
in a transparent manner by moving to the ``Einstein frame'' for the gravitons.
Performing a disformal (and conformal) transformation,\footnote{It was shown in Ref.~\cite{Domenech:2015hka}
that in cosmology a pure disformal transformation
is equivalent to rescaling the time coordinate.}
\begin{align}
\widetilde a &=\mpl^{-2}{\cal F}_T^{1/4}{\cal G}_T^{1/4}a,
\\
\D \widetilde t &=\mpl^{-2}{\cal F}_T^{3/4}{\cal G}_T^{-1/4}\D t,
\end{align}
the action~(\ref{action_hij}) can be recast into the standard form~\cite{Creminelli:2014wna},
\begin{align}
S_{h\,{\rm E}}^{(2)} =\frac{\mpl^2}{8}\int \D \widetilde t\D^3x \,\widetilde a^3
\left[(\partial_{\widetilde t}h_{ij})^2
-\widetilde a^{-2}(\Vec{\nabla}h_{ij})^2\right],
\end{align}
where the scale factor in this ``Einstein frame'' reads
\begin{align}
\widetilde a\sim (-\widetilde t\,)^{(n+1)/3}
\quad (-\infty < \widetilde t<0)
\end{align}
with $n>0$ for $\beta>1/2$.
Clearly, this is the scale factor of a contracting universe
where the tensor perturbations are effectively living.
This is the reason for the superhorizon growth of $h_{ij}$.
It is worth emphasizing that the frame we have moved to is the Einstein frame
only for the gravitons. This frame is not convenient for studying and interpreting
the background dynamics and the evolution of the curvature perturbation;
it was introduced just to understand the evolution of the tensor perturbations.

Growing tensor perturbations on superhorizon scales imply
that the anisotropic shear also grows.
Indeed, it can be seen that $\dot\beta_\pm$ in Sec.~\ref{subsec:aniso}
and $\dot h_{ij}$ share the same time dependence, $\propto(-t)^{-2\beta}$.
Nevertheless, this does not spoil the Genesis scenario because
the Hubble rate grows faster provided that $\alpha>-1/2$,
as discussed in Sec.~\ref{subsec:aniso}.

The power spectrum of the tensor perturbations
is dependent not only on ${\cal G}_T$ but also on ${\cal F}_T$,
and from Eq.~(\ref{behaviorFT}) one finds two distinct cases depending on whether
$2B'(Y_0)+Y_0B''(Y_0)$ vanishes or not.
Both cases yield the power spectrum of the form
\begin{align}
{\cal P}_h=A_Tk^{n_t}.
\end{align}
Since the explicit expression for $A_T$ is messy,
we do not give it here. Based on a concrete example we will evaluate $A_T$ in the next section.
To see the spectral index, let us first consider the case of $2B'(Y_0)+Y_0B''(Y_0)=0$.
In this case, we have ${\cal F}_T=2e^{-2\beta\lambda\phi}Y_0A'(Y_0)\propto (-t)^{2\beta}$.
It follows from Appendix~\ref{appendix1} that
the spectral index is dependent only on the parameter $\beta$ and is given by
\begin{align}
n_t=3-2|\nu|
\quad{\rm with}\quad \nu:=\frac{1}{2}-\beta,
\end{align}
where the constant mode is dominant for $\nu>0$, while
the would-be decaying mode grows for $\nu<0$.
The flat spectrum is obtained for $\beta= -1,\,2$.
In the case of $2B'(Y_0)+Y_0B''(Y_0)\neq 0$ we have ${\cal F}_T\propto (-t)^{2(\alpha+2\beta)}$,
so that $n_t$ is determined from the two parameters $\alpha$ and $\beta$ as
\begin{align}
n_t=3-2|\nu|
\quad{\rm with}\quad \nu:=\frac{1-2\beta}{2(\alpha+\beta+1)}.\label{tensor2}
\end{align}
The flat spectrum is obtained for $3\alpha + 5\beta+2=0$ and $3\alpha+\beta+4=0$,
though the latter case is not allowed under the conditions $\alpha+\beta>0$ and $\alpha>-1/2$.

In the previous study, we typically have blue spectra for tensor perturbations
in NEC-violating alternatives to inflation.
However, we have confirmed that
the spectra can also be flat and red, depending on the parameters,
in our variants of Galilean Genesis.

\subsection{Curvature Perturbation}\label{scalar-perturbation}
The quadratic action for the curvature perturbation
is~\cite{Kobayashi:2011nu} 
\begin{eqnarray}
S_\zeta^{(2)}=\int\D t\D^3x\,a^3\left[{\cal G}_S\dot\zeta^2+\frac{{\cal F}_S}{a^2}(\Vec{\nabla} \zeta)^2\right],
\end{eqnarray}
where 
\begin{eqnarray}
&& {\cal G}_S:=\frac{\Sigma {\cal G}_T^2}{\Theta^2}+3{\cal G}_T, \\
&& {\cal F}_S:=\frac{1}{a}\frac{\D}{\D t}\left(\frac{a{\cal G}_T^2}{\Theta}\right)-{\cal F}_T,
\end{eqnarray}
and
in the present class of Genesis models $\Sigma$ and $\Theta$ are given by
\begin{align}
\Sigma &\simeq e^{2(1+\alpha)\lambda \phi}Y_0 \hat\rho '(Y_0) \propto (-t)^{-2(\alpha +1)},
\\
\Theta &\simeq -e^{2\alpha\lambda\phi}Y_0\dot\phi g_3'
\notag \\ & \quad
-2e^{-2\beta\lambda\phi}HY_0(3A'+12Y_0A''+4Y^2_0A''')
\notag \\ & \quad
+e^{-2(1+\alpha+2\beta)\lambda\phi}H^2Y_0\dot\phi
\notag \\ & \quad \quad
\times (30B'+75Y_0B''+36Y^2_0B'''+4Y^3_0B^{(4)})
\notag \\ &
\propto (-t)^{-(2\alpha+1)}.
\end{align}
Thus, we have again two distinct cases and
if $\hat\rho '(Y_0)=0$ we find that
\begin{eqnarray}
{\cal G}_S \simeq 3{\cal G}_T \propto (-t)^{2\beta},
\end{eqnarray}
while if $\hat\rho '(Y_0)\neq 0$ we obtain 
 \begin{eqnarray}
{\cal G}_S \simeq \frac{\Sigma {\cal G}_T^2}{\Theta^2} \propto (-t)^{2(\alpha+2\beta)}.
\end{eqnarray}
Irrespective of whether ${\cal F}_T\propto (-t)^{2\beta}$ or $\propto (-t)^{2(\alpha+2\beta)}$,
we have
\begin{eqnarray}
{\cal F}_S \simeq \partial_t \left(\frac{{\cal G}_T^2}{\Theta}\right)-{\cal F}_T \propto (-t)^{2(\alpha+2\beta)}.
\end{eqnarray}

Now, from the formulas in Appendix~\ref{appendix1} it is easy to evaluate the
power spectrum of the curvature perturbation,
\begin{align}
{\cal P}_\zeta = A_S k^{n_s-1}.
\end{align}
Again, the explicit expression for $A_S$ turns out to be messy.
Therefore, $A_S$ will be evaluated through a concrete example in the next section
and here we focus only on the spectral index.
In the special case of $\hat \rho'(Y_0)=0$, we obtain the spectral index
\begin{align}
n_s-1=3-2|\nu|
\quad {\rm with}\quad
\nu:=\frac{1-2\beta}{2(\alpha+\beta+1)},
\end{align}
which shares the same expression as Eq.~(\ref{tensor2}).
The spectrum is therefore scale invariant for $3\alpha + 5\beta+2=0$.
In the general case of $\hat\rho'(Y_0)\neq 0$ we have
\begin{align}
n_s-1=3-2|\nu|
\quad {\rm with}\quad
\nu:=\frac{1}{2}-\alpha-2\beta,
\end{align}
and the spectrum is scale invariant for 
the parameters satisfying $\alpha+2\beta+1=0$ or $\alpha+2\beta-2=0$.

\begin{figure}[tbp]
  \begin{center}
  \includegraphics[keepaspectratio=true,width=85mm]{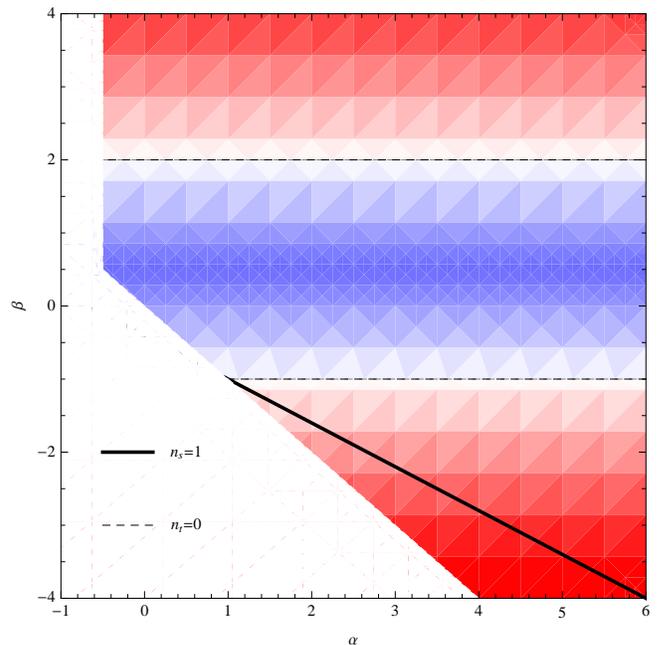}
  \end{center}
  \caption{Tensor and scalar tilts as functions of $\alpha$ and $\beta$
  in models with $2B'(Y_0)+Y_0B''(Y_0)=0$ and $\hat\rho'(Y_0)=0$, plotted
  in a viable parameter range, $\alpha+\beta>0$ and $\alpha>-1/2$.
  The thick solid line shows the parameters giving a scale-invariant spectrum of the curvature perturbation,
  $n_s=1$. The dashed lines correspond to scale-invariant tensor perturbations, $n_t=0$,
  and the red (blue) region represents the parameters for which the tensor spectrum is red (blue).
  For a nearly scale-invariant scalar spectrum, $n_s\approx 1$, only a red tensor spectrum is obtained
  in this class of models.
  }%
  \label{spectrum21.eps}
\end{figure}
\begin{figure}[tbp]
  \begin{center}
  \includegraphics[keepaspectratio=true,width=85mm]{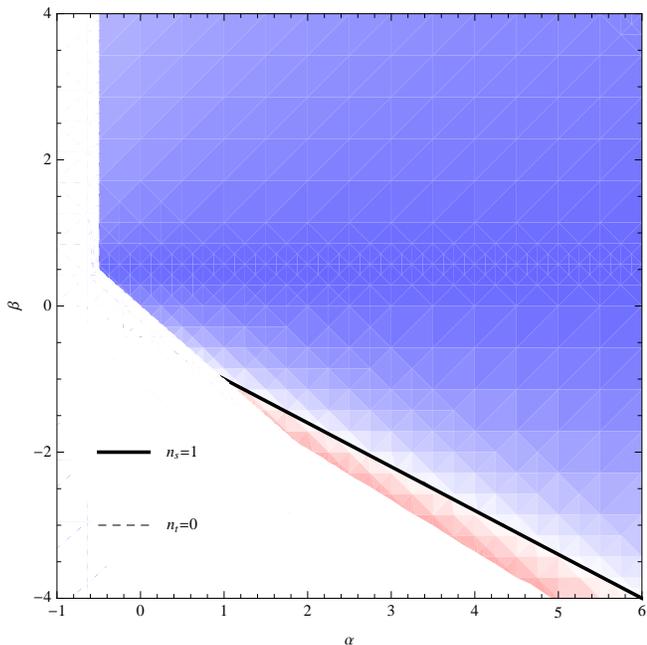}
  \end{center}
  \caption{Same as Fig.~\ref{spectrum21.eps}, but for models with
  $2B'(Y_0)+Y_0B''(Y_0)\neq 0$ and $\hat\rho'(Y_0)=0$.
  In this case, $n_t=n_s-1$, and hence
  the lines giving scale-invariant tensor and scalar spectra coincide.
  For $n_s=0.96$, the tensor tilt is given by $n_t=0.96-1=-0.04$.
  }%
  \label{spectrum22.eps}
\end{figure}
\begin{figure}[tbp]
  \begin{center}
  \includegraphics[keepaspectratio=true,width=85mm]{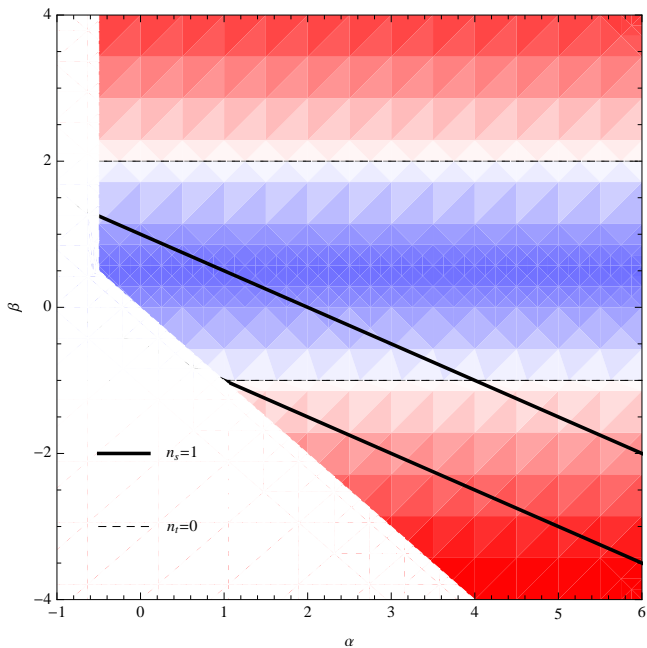}
  \end{center}
  \caption{Same as Fig.~\ref{spectrum21.eps}, but for models with
  	$2B'(Y_0)+Y_0B''(Y_0)=0$ and $\hat\rho'(Y_0)\neq 0$.
  	Both tensor and scalar spectra are scale invariant for $(\alpha,\beta)=(1,-1),\,(4,-1)$,
  	though the former is located at the edge of the viable parameter range.
  	For a nearly scale-invariant scalar spectrum, $n_s\approx 1$, both red and blue tensor spectra are possible
  	in this class of models.
  }%
  \label{spectrum11.eps}
\end{figure}
\begin{figure}[tbp]
  \begin{center}
  \includegraphics[keepaspectratio=true,width=85mm]{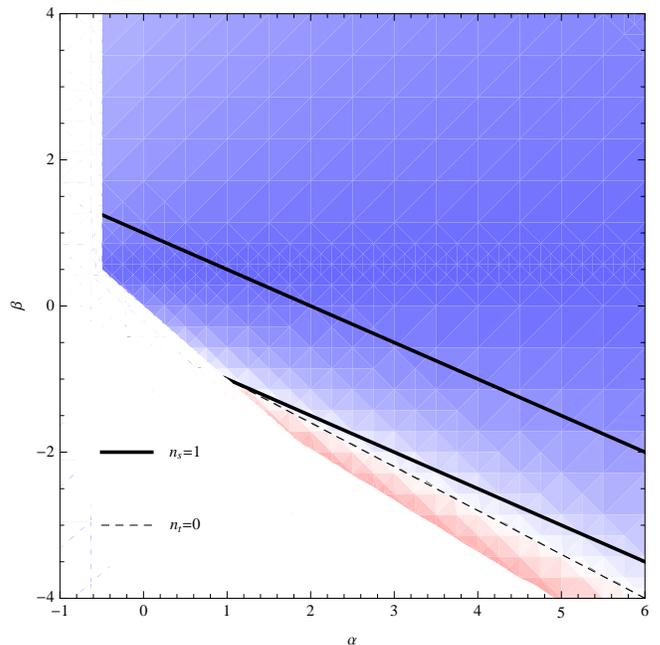}
  \end{center}
  \caption{Same as Fig.~\ref{spectrum21.eps}, but for models with
$2B'(Y_0)+Y_0B''(Y_0)\neq 0$ and $\hat\rho'(Y_0)\neq 0$.
  	For a nearly scale-invariant scalar spectrum, $n_s\approx 1$, the tensor spectrum is always blue
  	in this class of models.
  }%
  \label{spectrum12.eps}
\end{figure}

Having thus obtained the spectral indices $n_t$ and $n_s$, we summarize
the results in Figs.~\ref{spectrum21.eps}--\ref{spectrum12.eps}.
Of particular interest are the cases presented in Figs.~\ref{spectrum22.eps}
and~\ref{spectrum11.eps}.
In the former case both scalar and tensor perturbations have nearly scale-invariant
spectra for $3\alpha + 5\beta+2\simeq 0$,
while in the latter case this is possible for $\alpha\simeq 4$ and $\beta\simeq -1$.
In the other two cases, i.e., the cases given in Figs.~\ref{spectrum21.eps} and~\ref{spectrum12.eps},
the parameters leading to scale-invariant scalar and tensor perturbations
are on the boundaries of the viable parameter regions.

Before closing this section, let us comment on the stability of
the Genesis solutions. By requiring that ${\cal G}_T,\,{\cal F}_T,\,{\cal G}_S,\,{\cal F}_S>0$,
one can obtain a stable Genesis phase.
However, as shown in~\cite{Libanov:2016kfc,Kobayashi:2016xpl},
non-singular cosmological solutions in the Horndeski theory
are plagued with
gradient instabilities which occur at some moment in the entire expansion history,
provided that the integrals
\begin{align}
\int_{-\infty}^t a{\cal F}_T\D t'\quad
{\rm and}\quad
\int^{\infty}_t a{\cal F}_T\D t'
\end{align}
do not converge.
The Genesis models
with ${\cal F}_T\sim(-t)^n$ $(n\ge -1)$
satisfies the postulates
of this no-go theorem, and hence, even though a single genesis phase itself is stable,
gradient instability occurs eventually after the Genesis phase.
The models with ${\cal F}_T\sim(-t)^n$ $(n< -1)$
can evade the no-go theorem~\cite{Kobayashi:2016xpl},
but then the universe would be geodesically incomplete for gravitons~\cite{Creminelli:2016zwa}.
If one would prefer a geodesically complete universe for gravitons,
some new terms beyond
Horndeski must be introduced to avoid gradient
instabilities~\cite{Creminelli:2016zwa,Pirtskhalava:2014oc,Kobayashi:2015gga,Cai:2016thi}.

\section{An example}

As a concrete example, let us focus on the
case with $B(Y)=0$,
$\alpha = 4$, and $\beta=-1$,
which gives rise to exactly scale-invariant spectra for scalar and tensor perturbations.
Our example is given by
\begin{align}
g_2=-Y+\frac{Y^2}{\mu^4},
\quad
g_3=\frac{Y}{8\lambda \mu^4},\label{defg23}
\end{align}
and 
\begin{align}
A=M^2\left[\frac{Y}{\mu^4}-\left(\frac{Y}{\mu^4}\right)^2
+\frac{2}{5}\left(\frac{Y}{\mu^4}\right)^3\right],
\end{align}
where
$\mu$ and $M$ are parameters having dimension of mass, and
it follows from Eq.~(\ref{defg23}) that $Y_0=2\mu^4/3$.

One can solve the background equations to obtain
\begin{align}
H=\frac{5\sqrt{3}}{56}\frac{\mu^2}{\lambda M^2}e^{7\lambda\phi}\propto (-t)^{-7}.
\end{align}
It is straightforward to compute
\begin{align}
&{\cal G}_T=\frac{4}{9}M^2e^{2\lambda\phi},\quad
&&{\cal F}_T=\frac{4}{15}M^2e^{2\lambda\phi},
\notag \\
&{\cal G}_S=\frac{49}{72}\lambda^2M^4e^{-4\lambda\phi},
&&{\cal F}_S=\frac{70}{27}\lambda^2M^4e^{-4\lambda\phi},
\end{align}
which shows that this model is stable.
The primordial power spectra of tensor and curvature perturbations are given, respectively, by
\begin{align}
{\cal P}_h=\sqrt{\frac{5}{3}}\frac{10  \lambda^2 \mu^4}{\pi^2 M^2}
\simeq
1.3\times \frac{ \lambda^2 \mu^4}{ M^2},
\end{align}
and
\begin{align}
{\cal P}_\zeta\simeq 0.017\times (\lambda H_\ast)^{6/7}\left(\frac{\mu}{M}\right)^{16/7},
\end{align}
where $H_\ast$ is the Hubble parameter at the end of the genesis phase.
Note that the curvature perturbation grows on superhorizon scales and hence
${\cal P}_\zeta$ depends on the time when the Genesis phase ends,
while tensor perturbations do not.
The tensor-to-scalar ratio has a non-standard expression
(i.e., it does not depend on $n_t$ or the slow-roll parameter)
and reads
\begin{align}
r\sim 10^{-2}\times (\lambda H_\ast)^{-6/7} (\lambda \mu)^{12/7}(\lambda M)^{2/7},
\end{align}
which can be made sufficiently small by choosing the parameters.

One can improve the above model by introducing
slight deviations from $\alpha = 4$ and $\beta=-1$ to have $n_s\simeq 0.96$.
The lesson we learn from this example is that it is rather easy to
construct a stable model of Galilean Genesis
generating primordial curvature perturbations that are consistent with observations
and tensor perturbations that can be hopefully detected by future observations.

\section{Conclusions}

In this paper, we have proposed a variant of generalized Galilean Genesis
as a possible alternative to inflation.
A general Lagrangian for this new class of models has been constructed
within the Horndeski theory.
The Lagrangian has four functional degrees of freedom in addition to
two constant parameters, and includes the model studied in Ref.~\cite{Cai:2016gjd}
as a specific case.
We have confirmed that under certain conditions the background evolution of our Genesis models
leads to a stable, homogeneous and isotropic universe with flat spatial sections.
We have then calculated power spectra of primordial perturbations
and shown that
a variety of tensor and scalar spectral tilts can be obtained,
as summarized in Figs.~\ref{spectrum21.eps}--\ref{spectrum12.eps}.
In some cases, curvature/tensor perturbations grow on superhorizon scales
and for this reason the primordial amplitudes depend
not only on the functions in the Lagrangian but also
on the time when the Genesis phase ends.
It should be emphasized that
in spite of the gross violation of the null energy condition
the tensor spectrum can be (nearly) scale invariant,
though the consistency relation is still non-standard.

We have thus seen that in the Galilean Genesis scenario
both scalar and tensor power spectra can be nearly scale invariant
as in the standard inflationary scenario.
It is therefore crucial to evaluate the amount of non-Gaussianities
in the primordial curvature perturbations produced during the Genesis phase.
This point will be reported elsewhere.

\acknowledgments
This work was supported in part by the JSPS Research Fellowships for Young Scientists No.~15J04044 (S.N.)
and
the JSPS Grants-in-Aid for
Scientific Research No.~16H01102 and No.~16K17707 (T.K.).

\appendix

\section{Useful formulas for the power spectrum}\label{appendix1}

In this Appendix, we give some useful formulas for the power spectra of
cosmological perturbations in the case where the coefficients of kinetic and gradient terms
are of the power-law form.

Let us consider the quadratic action of the curvature perturbation $\zeta$ of the form
\begin{align}
S^{(2)}_\zeta =\int\D t\D^3x\left[C_1(-t)^{2p}\dot\zeta^2-C_2(-t)^{2q}(\Vec{\nabla}\zeta)^2\right],
\end{align}
where
$C_1$ and $C_2$ are positive constants
and the range of the time coordinate is $-\infty<t<0$.
It is convenient to introduce a new time coordinate defined as
\begin{align}
-y:=\frac{C_2^{1/2}}{C_1^{1/2}}\frac{(-t)^{1-p+q}}{1-p+q}.
\end{align}
Assuming that $1-p+q>0$,
the $y$ coordinate also ranges from $-\infty$ to $0$.
The canonical variable is
\begin{align}
u:=\sqrt{2}(C_1C_2)^{1/4}(-t)^{(p+q)/2}\zeta,
\end{align}
in terms of which the action is written as
\begin{align}
S^{(2)}=\frac{1}{2}\int\D y\D^3x\left[
\left(\frac{\partial u}{\partial y}\right)^2-(\Vec{\nabla}u)^2+\frac{\nu^2-1/4}{y^2}u^2\right],
\end{align}
where
\begin{align}
\nu:=\frac{1-2p}{2(1-p+q)}.
\end{align}
This is the familiar form of the action for the Sasaki-Mukhanov variable.

The positive frequency solution in the Fourier space reads
\begin{align}
u_k=\frac{\sqrt{\pi}}{2}\sqrt{-y}H_{\nu}^{(1)}(-k y).
\end{align}
Horizon crossing occurs when $|ky|\sim 1$,
and then for $|ky|\ll 1$ we find
\begin{align}
|\zeta_k|&\simeq \frac{2^{|\nu|-3/2}}{\pi}\frac{\sqrt{\pi}}{(C_1C_2)^{1/4}}
\left[
(1-p+q)\frac{C_1^{1/2}}{C_2^{1/2}}(-y)\right]^{\nu-1/2} \notag \\
&
\quad \times k^{-|\nu|}(-y)^{1/2-|\nu|}
\notag \\
&
\propto k^{-|\nu|}|y|^{\nu-|\nu|}.
\end{align}
This implies that as in the usual case
the constant mode dominates
for $\nu>0$, while the curvature perturbation grows on superhorizon scales for $\nu<0$.
In both cases, we obtain a scale-invariant spectrum for $|\nu|=3/2$.
The concrete expression for the power spectrum
${\cal P}_\zeta$ (evaluated at some $y$) is given by
\begin{align}
{\cal P}_\zeta = \left[\frac{2^{|\nu|-3/2}\Gamma(|\nu|)}{\Gamma(3/2)}\right]^2\frac{k^{3-2|\nu|}}{8\pi^2}
\frac{C_1^{\nu-1}}{C_2^{\nu}} \nonumber \\
\times (1-p+q)^{2\nu-1}|y|^{2(\nu-|\nu|)}.
\end{align}

For the tensor perturbations whose action is given by
\begin{align}
S_h^{(2)}=\frac{1}{8}\int\D t\D^3x\left[C_1(-t)^{2p}\dot h_{ij}^2-C_2(-t)^{2q}(\Vec{\nabla}h_{ij})^2\right],
\end{align}
The calculation is essentially the same as that demonstrated for $\zeta$,
and the power spectrum is given by
\begin{align}
{\cal P}_h=16\left[\frac{2^{|\nu|-3/2}\Gamma(|\nu|)}{\Gamma(3/2)}\right]^2\frac{k^{3-2|\nu|}}{8\pi^2}
\frac{C_1^{\nu-1}}{C_2^{\nu}} \nonumber \\
\times(1-p+q)^{2\nu-1}|y|^{2(\nu-|\nu|)}.
\end{align}




\end{document}